\documentstyle[prl,twocolumn,aps,psbox]{revtex}
\def\apj #1 #2 #3 {#1, ApJ, {\bf #2}, #3}
\def\apjl #1 #2 #3 {#1, ApJ, {\bf #2}, L#3}
\def\apjs #1 #2 #3 {#1, ApJS, {\bf #2}, #3}
\def\aap  #1 #2 #3 {#1, A\&A, {\bf #2}, #3}
\def\mnras #1 #2 #3 {#1, MNRAS, {\bf #2}, #3}
\def\pra #1 #2 #3 {#1, Phys.~Rev.~A., {\bf #2}, #3}
\def\prb #1 #2 #3 {#1, Phys.~Rev.~B., {\bf #2}, #3}
\def\prc #1 #2 #3 {#1, Phys.~Rev.~C., {\bf #2}, #3}
\def\prd #1 #2 #3 {#1, Phys.~Rev.~D., {\bf #2}, #3}
\def\pre #1 #2 #3 {#1, Phys.~Rev.~E., {\bf #2}, #3}
\def\prl #1 #2 #3 {#1, Phys.~Rev.~Lett., {\bf #2}, #3}
\def\plb #1 #2 #3 {#1, Phys.~Lett.~B., {\bf #2}, #3}
\def\science #1 #2 #3 {#1, Science., {\bf #2}, #3}
\def\nature #1 #2 #3 {#1, Nature., {\bf #2}, #3}
\def\nphysa #1 #2 #3 {#1, Nucl.~Phys.~A., {\bf #2}, #3}
\def\nphysb #1 #2 #3 {#1, Nucl.~Phys.~B., {\bf #2}, #3}
\def\nphysbs #1 #2 #3 {#1, Nucl.~Phys.~B.~Suppl., {\bf #2}, #3}

\def\h#1{\hbox{${}^{#1}$H}}
\def\he#1{\hbox{${}^{#1}$He}}

\def\dovh{\hbox{D/H}}

\def\yp{\hbox{\hbox{$Y_p$}}}

\def\h502{\hbox{$ h^{2}_{50}$}}

\def\fun#1#2{\lower3.6pt\vbox{\baselineskip0pt\lineskip.9pt
  \ialign{$\mathsurround=0pt#1\hfil##\hfil$\crcr#2\crcr\sim\crcr}}}
\begin{document}
\draft
\title{Constraints on Cosmic Quintessence and Quintessential Inflation}
\author{
 M. Yahiro }
\address{
Department of Physics and Earth Sciences, University of the Ryukyus,
Nishihara-chou, Okinawa 903-0213, Japan }
\author{G. J. Mathews}
\address{Center for Astrophysics, 
Department of Physics, University of Notre Dame, Notre Dame, IN 46556 }
\author{
 K. Ichiki$^{1,2}$, T. Kajino$^{1,2,3}$ and  M. Orito$^1$
}
\address{
$^1$National Astronomical Observatory, 2-21-1, Osawa, Mitaka, Tokyo 181-8588, 
Japan}
\address{
$^2$University of Tokyo, Department of Astronomy, 7-3-1
Hongo, Bunkyo-ku, Tokyo 113-0033, Japan }
\address{
$^3$Graduate University for Advanced Studies, Dept.
of Astronomical Science, 2-21-1, Osawa, Mitaka, Tokyo 181-8588, Japan}
\date{\today}
\maketitle
\begin{abstract}
Recently, attempts have been made to understand the apparent near
coincidence of the present dark energy and matter energy in terms of
a dynamical attractor-like solution for 
the evolution of a ``quintessence" scalar field.  
In these models the field couples with the dominant
constituent and only acts like a cosmological constant 
after the onset of the matter dominated epoch.  
A generic feature of such solutions, however, is the possibility of
significant energy density in the scalar field during the
radiation dominated epoch.  This possibility is even greater if the
quintessence  field begins in a kinetic-dominated  regime
generated at the end of "quintessential
inflation."  As such, these models can affect,
and therefore be constrained by, primordial nucleosynthesis
and the epoch of photon decoupling.
Here we analyze one popular form for 
the quintessence field (with and without a supergravity correction)
 and  quantify constraints on
the allowed initial conditions and parameters for the effective potential.
We also deduce constraints on the epoch of matter creation at the end
of quintessential inflation.
\end{abstract}
PACS numbers; 98.80.Cq, 98.65.Dx, 98.70.Vc
%

\section{INTRODUCTION}
Recent observations \cite{garnavich,perlmutter} 
of Type Ia supernovae at intermediate
redshift, along with complementary observational constraints
at low  and intermediate redshift, as well as 
the power spectrum of the cosmic microwave background all
indicate  \cite{wang}
 that the universe may be presently accelerating due to
the influence of a dominant dark energy with a negative pressure.
The simplest  interpretation
of this dark energy is a cosmological constant 
for which the equation of
state is $\omega \equiv P/\rho  = -1$.  
A second possibility is derived from the so-called ``quintessence" models.  
In this case the dark   energy is the result of a scalar field 
$Q$ slowly evolving along an effective
potential $V(Q)$.  The equation of state 
is negative $-1 \le w_Q \le 0$,
but not necessarily constant. 

Introducing either of these paradigms, 
however, leads inevitably to two nagging questions.
One is a fine tuning problem as to 
why the  present dark  energy is so small compared to 
the natural scales of high-energy physics.
The second  is a cosmic coincidence problem 
as to why the universe has conspired
to produce nearly equivalent energy content in matter and dark   energy
at the present time.    

  Attempts have been made 
\cite{wetterich,zlatev,chiba,armendariz1,armendariz2,steinhardt99,brax} 
to reformulate this quandary
by introducing  specific forms of the quintessence effective 
potential whereby a 
tracker field $Q$ evolves
according to an attractor-like solution to the equations
of motion.  That is, for a wide variety of initial conditions,
the solutions for $Q$ and $\dot Q$  rapidly approach a common
evolutionary track. The nice feature of these solutions
is that they lead naturally to a cross over from an earlier 
radiation-dominated solution to one in which the universe 
is dominated by a small dark  energy at late times.
Another interesting possible feature is that such models might
also naturally arise during matter creation at the
end of an  earlier  ''quintessential`` inflationary epoch 
\cite{vilenkin}.
In this case,  the $Q$ field emerges in a 
kinetic-dominated regime at energy densities above the tracker 
solution.

It is not yet clear, however, \cite{kolda99,copeland} that these models
have altogether solved the fine-tuning and cosmic-coincidence problems.
Nevertheless, several such tracker fields 
have been proposed \cite{brax}.  Although there is some 
difficulty in aligning quintessence models and string theory
\cite{hellerman}, the form for the effective potentials
may at least be suggested by particle physics models
with dynamical symmetry breaking, by nonperturbative effects
\cite{zlatev}, by generic kinetic terms "$k$-essence"  in an  effective
action describing the moduli and massless degrees of freedom in string and
supergravity theories \cite{chiba,armendariz1,armendariz2},  
or by static and moving branes in
a dilaton gravity background \cite{chenlin}.  
 
A general feature of all such solutions,
however, is that at least the possibility exists for a significant
contribution of the $Q$ field to the total energy density during the
radiation-dominated or photon decoupling  epochs as well as
the present matter-dominated epoch.  The
yields of primordial nucleosynthesis  and the power spectrum
of the CMB are strongly affected by
the background energy density and universal expansion rate.
Therefore,
the possibility exists to utilize primordial nucleosynthesis
and the CMB power spectrum
to constrain otherwise viable quintessence or  $k$-essence models.

Observational constraints on such quintessence models have been
of considerable recent interest.  For example, in \cite{bean}
the constraints on an  $M$-theory motivated 
exponential form with a quadratic
prefactor  \cite{as} for the quintessence effective
potential were considered.  In this model, the 
$Q$ field closely follows the background field.
In \cite{chen} a study was made of the effect 
on primordial nucleosynthesis of 
extended quintessence with nonminimal coupling.
In such models the quintessence field couples with the scalar curvature.
It can then act to slow the expansion
and affect the yields of primordial nucleosynthesis.  In the present work, we 
consider the power-law effective potential and its supergravity-corrected form
as described below.  We construct a detailed mapping of the allowed
parameter space  for these quintessence models.  We also consider 
constraints these considerations place on the epoch of matter
generation at the end of quintessential inflation.

\section{Quintessence Field}
 A variety of quintessence \cite{brax} effective potentials 
or $k$-essence effective actions \cite{chiba,armendariz1,armendariz2}
can be found in the literature.  
  In this paper, however,  we  will not consider $k$-essence
effective actions as we have found that the attractor-like solution 
is limited to a prohibitively  small range of parameters of the effective 
Lagrangian.  We will, however, consider kinetic dominated quintessence models.
Here, we 
concentrate on what is a frequently invoked
form for the effective potential of
the tracker  field, i.e.  an inverse power law such as
originally analyzed by Ratra and Peebles \cite{ratra},
\begin{equation}
V(Q) = M^{(4  +  \alpha)} Q^{-\alpha}~~,
\end{equation}
where, $M$  and $\alpha$ are parameters.
The parameter $M$ in these potentials is fixed by the
condition that $\Omega_Q = 0.7$ at present.
Therefore, 
\begin{equation}
\rho_Q(0) = 0.7 \rho_c(0) = 5.7 h^2 \times 10^{-47}$ GeV$^4
\label{rhoq0}
\end{equation}
 and
\begin{equation}
M  \approx  \biggl(\rho_Q(0)Q^\alpha\biggr)^{1/(\alpha +4)}~~.
\label{mfix}
\end{equation}
If $Q$ is presently near the Planck mass and $\alpha$ is too small
(say $\alpha ^<_\sim 2$), this 
implies  a small value \cite{zlatev} for $M$  which in a sense
reintroduces the fine tuning problem.  

 We will also consider a modified  form of $V(Q)$ as proposed
by \cite{sugraref} based upon the condition that
the quintessence fields be part of supergravity models.
The rewriting of the effective potential in supergravity
depends upon choices of the K\"ahler potential \cite{copeland}.
The flat K\"ahler potential yields an extra factor of
$exp{\{3Q^2/2m_{pl}^2\}}$ \cite{sugraref}.  This comes about
by imposing the condition that the expectation 
value of the superpotential vanishes.
The Ratra potential thus becomes 
\begin{equation}
V_{SUGRA}(Q) \rightarrow  
M^{(4 + \alpha)} Q^{-\alpha} \times exp{(3Q^2/2m_{pl}^2)}~~,
\end{equation}
where the exponential correction becomes largest near the present time
as $Q \rightarrow  m_{pl}$.  
This  supergravity motivated  effective potential is known as
the  SUGRA potential.  The fact that this potential 
has a minimum for $Q = \sqrt{\alpha/3} m_{pl}$ changes the dynamics.
It causes the present value of $w_Q$ to evolve to a
cosmological constant ($w_Q \approx -1$) much quicker than for the bare
power-law potential \cite{brax}.

The quintessence field $Q$ obeys the equation of motion
\begin{equation}
\ddot Q  + 3 H \dot Q  + d V/dQ  = 0~~,
\label{qddot}
\end{equation}
where the Hubble parameter $H$ is given from the solution to the
Friedmann equation,
\begin{equation}
H^2  = \biggl({\dot a \over a}\biggr)^2  =  {1 \over m_{pl}^2} 
(\rho_B +  \rho_Q)~~,
\label{friedmann}
\end{equation}
where $m_{pl} = (8\pi G/3)^{-1/2} = 4.2 \times 10^{18}$ GeV,
$\rho_B$ is the energy density in background radiation and matter,
and  $a$ is the cosmic scale factor.

For the simple inverse power-law potential,
it can be shown \cite{ratra} that the tracker solution maintains the condition
\begin{equation}
d^2 V/d Q^2 = (9/2)(1 - \omega_Q^2)[(\alpha + 1)/\alpha]H^2~~.
\end{equation}
As the $Q$ field evolves, its contribution to the energy density
is given by
\begin{equation}
\rho_Q =  {\dot Q^2 \over 2}   +  V(Q)~~,
\end{equation}
which may or may not be comparable to the energy density in radiation
during the nucleosynthesis or photon-decoupling  epoch depending upon the
parameters and  initial conditions.

The quintessence initial conditions are  probably set in
place near the inflation epoch.  By the time of the 
big bang nucleosynthesis (BBN) epoch,
many of the possible initial conditions will have
already achieved the tracker solution.
However, for initial conditions sufficiently removed
from the tracker solution, it is quite possible that 
the tracker solution has not yet been  achieved by the time of BBN.  
Such possibilities are illustrated schematically
in Figure \ref{fig:1}.  These correspond
to cases in which the initial energy density falls above (curve $A$),  
near (curve $B$) or below (curve $C$) the tracker solution.

\section{Nucleosynthesis Constraint}
\subsection{Quintessence and BBN}
\begin{figure}
\psbox[xsize=3.3in]{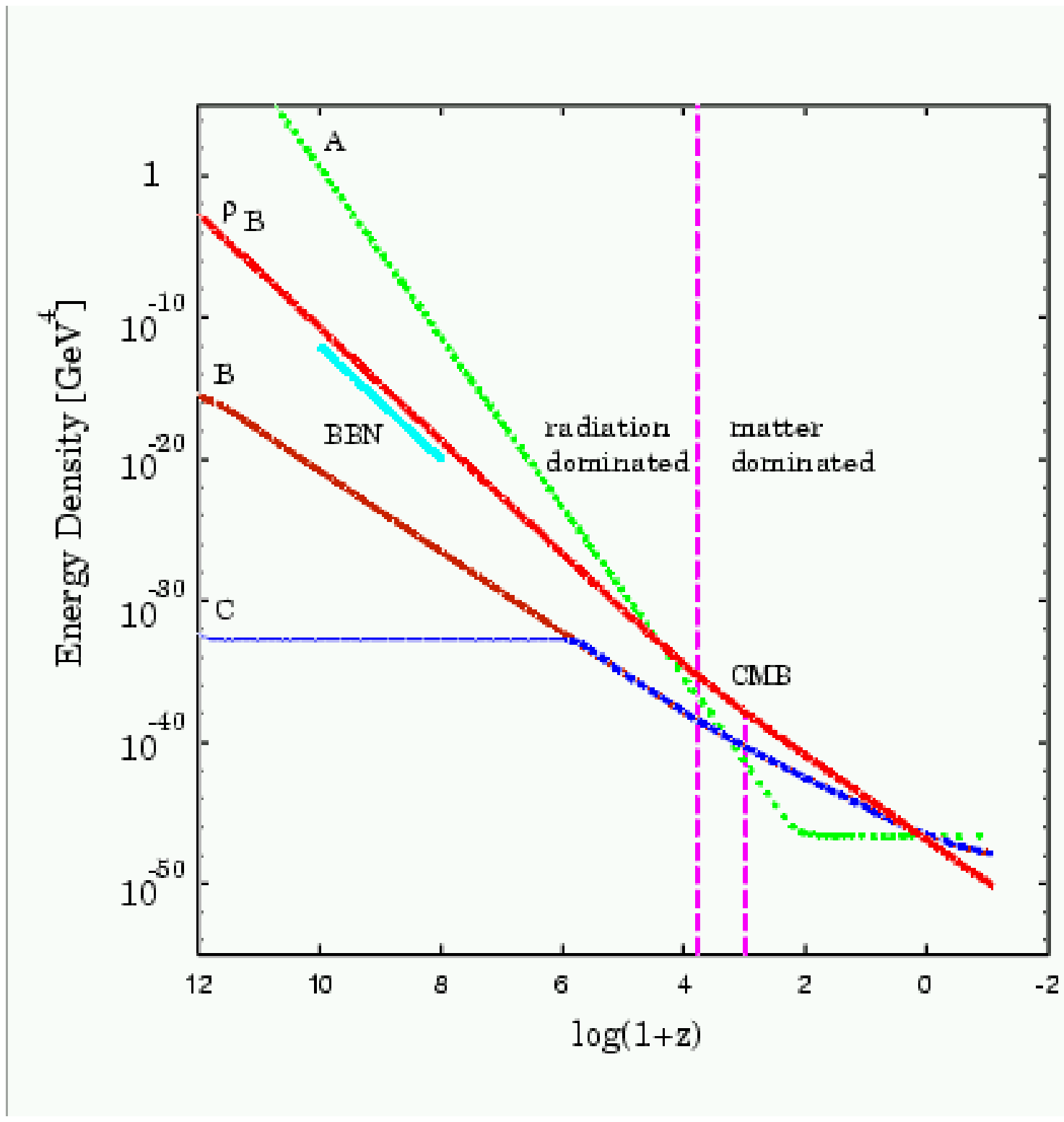}
\caption{Examples of the evolution of energy density in  $\rho_Q$ and 
the background fields $\rho_B$ as a function of redshift for an
inverse power-law effective potential with $\alpha = 5$.  
The BBN epoch is indicated by a short 
line segment.  The location of the
transition from radiation to matter
domination (for $\Omega_M = 0.3$), and the CMB epoch, 
are also indicated.  Curves
$A$  and $C$ illustrate cases in which the $Q$ field does not  
achieve the tracker solution until well after the BBN epoch.
Curve $A$ is excluded, curve $C$ is not.
Curve $B$ is an example of a tracker solution which is allowed
by the nucleosynthesis constraints.}
\protect
\label{fig:1}
\end{figure}

There are several paradigms of possible interest for constraint
by primordial nucleosynthesis.   These depend upon the initial
values for the energy density in the $Q$ field.
In any of these cases, the energy density in $\rho_Q$
can be constrained by the ratio $\rho_Q/\rho_B$
during the BBN epoch at 
$ 0.01 ^<_\sim T ^<_\sim 1$ MeV,
$10^{8}~^<_\sim z~^<_\sim 10^{10}$ as shown in Figure \ref{fig:1}.

If the initial conditions are sufficiently close to the
tracker solution $\bar \rho_Q$ before nucleosynthesis, 
then $\rho_Q \sim \bar \rho_Q$ during nucleosynthesis
and the tracker solution specifies the energy density in 
the $Q$ field.  
This situation is similar to curve $B$ in Figure 1. 
 This may be the most likely scenario. 
Along the tracker solution, $\rho_Q$
diminishes in a slightly different 
way than the radiation-dominant background energy density. 
For example, 
as long as $\rho_Q << \rho_B$,
the $Q$-field  decays as
\begin{equation}
 \rho_Q \propto  a^{-3(1+w_Q)}~~,
\label{qdecay}
\end{equation}
with  
\begin{equation}
w_Q = (\alpha w_B - 2)/(\alpha+2) < w_B~~.
\label{wq}
\end{equation}
The equation of state $w_Q$ is  only equal
to the background equation of state
$w_B$ in the limit $\alpha \to \infty$.  This is 
 equivalent to the exponential
potential.  Nevertheless, the tracker solution does not deviate much
from $\rho_B$, even at high redshift for most values  of
$\alpha$ considered here.
Hence, one can characterize the nucleosynthesis results
by the (nearly constant)  ratio  $\rho_Q/\rho_B$
during the BBN epoch.

If the energy density in the tracker solution
is close to the background energy density,
the nucleosynthesis will be affected by
the increased expansion rate
from the increased total energy density. Such a situation occurs for
large values of the power-law exponent $\alpha$.

A second possibility is
that the energy density  $\rho_Q$ could exceed the tracker solution
and be comparable to or greater than the background energy density
during primordial nucleosynthesis.  
This situation is something like curve $A$ in Figure \ref{fig:1}.
The kinetic energy in the $Q$ field dominates over
the potential energy contribution to $\rho_Q$ and  
$w_Q = +1$ so that the kinetic energy density  
diminishes as $a^{-6}$.
In this case
 there could be a significant contribution from $\rho_Q$
during nucleosynthesis as the $Q$ field approaches the tracker solution.  
The strongest constraints on this case would arise when
$\rho_Q$ is comparable to the background energy density  near the time of
the weak-reaction freese out, while the later nuclear-reaction epoch 
might be unaffected.  
This case $A$ is particularly interesting as this kinetic-dominated evolution 
could be generated by an earlier quintessential inflation epoch
\cite{vilenkin} as described below. 

A final possibility might occur if the $Q$ field approaches the
tracker solution from below as indicated by curve $C$ in Figure \ref{fig:1}.
In this case, the tracker solution may be achieved   
after the the BBN epoch so that a small $\rho_Q$ during BBN
is easily guaranteed.  In such models, however, the ultimate
tracker curve  might have
a large energy density at a later epoch which could be constrained
by the later CMB epoch as described below.

\subsection{Light-Element Constraints}
 The primordial light-element abundances
deduced from observations have been reviewed by a number
of recent papers \cite{osw99,nollett00,tytler00,steigman00}.
There are several outstanding uncertainties.  For primordial helium
there is an uncertainty due to the fact that the deduced abundances tend
to reside in one of  two possible values.
Hence, for \he4 we adopt a rather conservative range:
         $$ 0.226 \leq \yp \leq 0.247.  $$
For deuterium there is a similar possibility for either a
high or low value.  Here, however, we adopt the generally accepted
\cite{nollett00,tytler00}
low values of 
        $$ 2.9 \times 10^{-5} \leq \dovh \leq 4.0 \times 10^{-5}.  $$

The primordial lithium  abundance can be inferred from the
surface abundances of old halo stars.
There is, however, some uncertainty from the possibility
that old halo stars may have gradually depleted their primordial
lithium.  Because of this we do not consider primordial lithium
in the discussions here.  The adoption of the low deuterium abundance
forces the BBN primordial helium to be near its upper limit.

Adding energy density from the $Q$ field tends to increase
the universal expansion rate.  Consequently, the weak reaction rates
 freeze out at a higher temperature $T_w$.  
This fixes the neutron to proton ratio ($n/p \approx \exp[{(m_p -m_n)/T_w]}$)
at a larger value.  Since most of the free neutrons are converted into
$\he4$, the primordial helium production  is increased.
Also, since the epoch of nuclear reactions is shortened, the
efficiency of burning deuterium into helium is diminished
and the deuterium abundance also  increases.
Hence, very little excess 
energy density from the $Q$ field is allowed.

\begin{figure}
\psbox[xsize=3.3in]{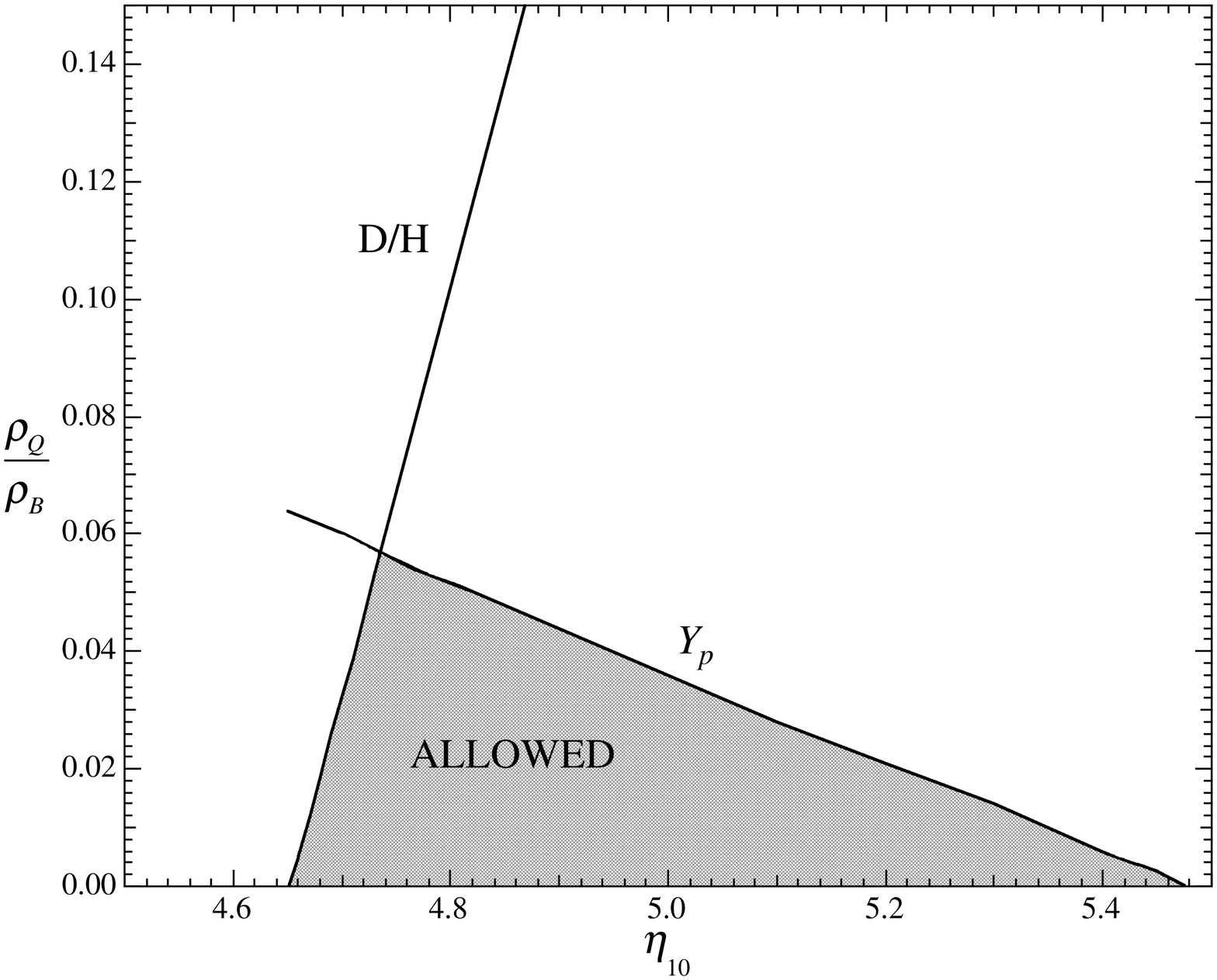}
\caption{Constraints on the ratio of the energy density in the quintessence
field to the background energy density  $\rho_Q/\rho_B$  (at $T = 1$ MeV)
from the primordial abundances as indicated. The allowed region corresponds
to $Y_p \le 0.247$ and D/H $\le 4.0\times 10^{-5}$.}
\protect\label{fig:2}
\end{figure}

Figure \ref{fig:2} summarizes allowed values of $\rho_Q/\rho_B$ at 
$T=$1 MeV  ($z \approx 10^{10}$) 
 based upon the nucleosynthesis constraints.
The region to the right of the curve labeled $D/H$ corresponds
to models in which the primordial deuterium constraint is satisfied,
$D/H \le 4.0 \times 10^{-5}$.  The region below the line labeled
$Y_p$ corresponds to $Y_p \le 0.247$.  The hatched region summarizes
allowed values for the energy density in the quintessence field
during the nucleosynthesis epoch.

This figure is similar to that obtained in 
Freese et al. \cite{freese}
based upon the light-element constraints available at that time.
They similarly considered dark   energy densities 
which scale proportionally to the background radiation energy density.
However, in their study a somewhat larger range of
possible dark   energies was allowed due to the larger uncertainties
in the observed primordial abundances and the neutron half life
at that time.

In the present work we deduce an absolute
upper limit of 5.6\%  of the background radiation energy density
allowed in the quintessence field.  This maximum contribution
is only allowed  for $\eta_{10}  \approx  4.75$ or $\Omega_b h^2
\approx  0.017$.   
A smaller contribution is allowed for
other values of $\eta_{10}$.  Indeed, this optimum $\eta_{10}$ value is
$4 \sigma$ 
less than the value implied by the cosmic deuterium abundance
\cite{nollett00,tytler00} 
$\Omega_b h^2 = 0.020 \pm 0.001~(1 \sigma)$ ($\eta_{10} = 
5.46 \pm 0.27$). 
The most recent independent determinations of $\Omega_b h^2$
from  high-resolution
measurements of the power spectrum of fluctuations in the 
cosmic microwave
background favor a value even higher.
Both the BOOMERANG  \cite{boom} and DASI  \cite{dasi}  data  sets imply
$\Omega_b h^2 = 0.022^{+ 0.004}_{-0.003} ~(1 \sigma)$ ($\eta_{10} =
6.00^{+ 1.10}_{-0.81}$).   The deuterium and CMB constraints together
demand that $\eta_{10} \ge 5.19$ which would limit the 
allowed contribution from the $Q$ field to $\le 2\%$ of the background energy
density.

The most restrictive CMB constraint  on $\Omega_b h^2$ derives from
demanding a flat universe ($\Omega_{tot} =  1.0$),
and marginalizing the likelihood function
over all parameters with assumed Gaussian
priors \cite{boom} based upon 
measurements of large-scale structure and Type Ia supernovae.  
 This gives $\Omega_b h^2 = 0.023 \pm 0.003~(1 \sigma)$.
If one adopts this  as a most extreme case, then
  $\Omega_b h^2 \ge 0.020$. This
would correspond to  $\eta_{10} \ge 5.46$.
From Figure 2 this would imply a much more stringent constraint 
that only about 0.1\% of the background energy density could 
be contributed by the $Q$ field.
Of course, this is only a $1\sigma$ constraint, and
it is questionable as to whether the upper limit
to $Y_p$ is well enough established to rule out a contribution
to the energy density at the 0.1\% level.
For the remainder of this paper we will
adopt the more conservative constraint of 5.6\%.  Nevertheless,
it is of interest to explore how the quintessence parameters
allowed by BBN might improve should
the constraints from BBN ever be so tightly 
defined.   Therefore, we will consider
 0.1\% as a conceivable limit that demonstrates the sensitivity to
BBN.   Even the most conservative  5.6\% limit  adopted here 
corresponds to only about half of the energy density allowed
in \cite{freese} for 3 neutrino flavors.

\section{Equation of State Constraint}

   Another constraint on allowed parameters for the quintessence field
derives from the simple requirement that the $Q$ field behaves like
dark  energy during the present matter-dominated epoch.
Stated another way, the equation of state should be sufficiently negative, 
i.e. $w_Q \equiv P_Q/\rho_Q < 0$ by the present time.

 In general, $w_Q$ is a time-dependent
quantity.  The energy density and pressure in the $Q$ field can be
written $\rho_Q = {1 \over 2} \dot Q^2 + V(Q)$ and
$P_Q = {1 \over 2} \dot Q^2 - V(Q)$.
This gives,
\begin{equation}
w_Q  = 1 - 2 V(Q)/\rho_Q~~,
\label{wqnum}
\end{equation}
where the time dependence derives from the evolution of $V(Q)$ and
$\rho(Q)$.  A comprehensive study of the observational constraints
on the present value of $w_Q$ has been recently summarized
by Wang et al. \cite{wang}.  We adopt the same constraints deduced
in that paper.  Mindful of systematic errors, they adopted a 
most conservative approach based upon progressively less reliable data sets.
Using the most reliable current low-redshift and microwave background
measurements, they deduce limits of $-1 < w_Q < -0.2$ at the 2$\sigma$
level.  Factoring in the constraint from Type Ia supernovae reduces
the range for the equation of state to $-1 < w_Q < -0.4$.  This
range derives from a concordance analysis of
models consistent with each observational
constraint at the $2\sigma$ level or better. 
A combined maximum likelihood analysis suggests
a smaller range of $-0.8 < w_Q < -0.6$  for quintessence models 
which follow the tracker solution,
though $w_Q \approx  -1$ is still allowed in models with nearly a constant
dark   energy. In what follows, we also invoke these same three
possible limit ranges for the present value of $w_Q$.

As we shall see,  these  limits   place a 
strong constraint on the  bare Ratra 
power-law potential for almost any value of $\alpha$.
 However,  the SUGRA corrected form of $V(Q)$ 
is only slightly constrained.

\section{CMB Constraint}

  A third constraint on the quintessence field arises from its effect on
the epoch of photon decoupling.    There are two effects
to be considered. 

\subsection{Look-back effect}
One is the effect of the $Q$ field on the observed microwave background 
in the case where the energy density in the quintessence field
is negligible during photon decoupling.  This has  been considered in \cite{brax}.  
In this case the effect of the dark   energy is to modify the
angular distance-redshift relation \cite{hu}.  The existence of dark   energy
during the look-back time to the surface of last scattering shifts the 
acoustic peaks in the CMB power spectrum to smaller angular scales
and larger $l$-values.
The amplitude of the first acoustic peak in the power spectrum also increases,
but not as much for quintessence  models  as for a
true cosmological constant $\Lambda$.  The basic features of the observed
power spectrum \cite{boom} can be fit \cite{brax} with either
of the quintessence potentials considered here.
Indeed, dark   energy is required to reproduce both the observed power spectrum
and the Type Ia supernova data.
For our purposes, this look-back  constraint
 is already satisfied by demanding that 
$\Omega_Q = 0.7$ at the present time
through Eqs.~(\ref{rhoq0}) and (\ref{mfix}).   

\subsection{Energy in $Q$ field}
 
There is, however, an additional possible
effect of the $Q$ field on the CMB which we also use as a constraint.
If the energy density in the $Q$ field is a significant fraction of the 
background energy during the epoch of
photon decoupling, it can increase
the expansion rate.  Increasing the expansion rate can also 
 push the $l$ values
for the acoustic peaks in the spectrum to larger values and increase  
their amplitude \cite{hu}.  Such an effect of an increased expansion
rate has been considered by a number of authors  in various  
contexts  \cite{bean,hu,hannestad}.
For our purposes, we adopt the constraint deduced by
\cite{bean} based upon the latest CMB sky maps of
the Boomerang \cite{boom} and DASI \cite{dasi} collaborations that
the density in the $Q$ field can not exceed $\Omega_Q \le 0.39$
during the epoch of photon decoupling.  This implies
a maximum allowed contribution of the $Q$ field  
during photon decoupling of $\rho_Q/\rho_B = \Omega_Q/(1-\Omega_Q)~^<_\sim 0.64$.
Note, however,
that the $Q$ field behaves differently near photon  decoupling than
during the BBN epoch (cf. Fig. \ref{fig:1}). After 
the transition from a radiation-dominated to a matter-dominated
universe, the $Q$ field (now coupled to the matter field)
can contribute a much larger fraction
of the background energy density (e.g. Curve $B$ of Fig. \ref{fig:1})
than in the BBN epoch. Hence,
the constraint on $\rho_Q$ field from 
BBN must be considered separately
from the constraint at the CMB epoch.   

\section{Quintessential Inflation}

Another possible constraint arises  if the kinetic term 
dominates at an early stage (e.g. Curve $A$ of Fig. \ref{fig:1}).
  In this case $\rho_Q \approx \dot Q^2/2$ 
and $\rho_Q$ decreases with scale factor as $a^{-6}$. 
At very early times this kinetic regime can be produced by  
so-called "quintessential inflation" \cite{vilenkin}.
In this paradigm entropy in the matter fields comes from gravitational
particle production at the end of inflation.  
The universe is presumed to exit from inflation directly 
into a  kinetic-dominated quintessence regime during which 
the matter is generated.
An unavoidable consequence of this process, however,
is the generation of gravitational waves along with matter and the 
quanta of the quintessence field 
\cite{vilenkin,riazuelo,giovannini1,giovannini2,ford}
at the end of inflation.

\subsection{Energy in Quanta and Gravity Waves}
Particle creation at the end of inflation has been studied by the
standard methods of quantum field theory in curved space-time \cite{ford}.
The energy density in created particles is just 
\cite{vilenkin,giovannini1,giovannini2,ford}
\begin{equation}
\rho_B \approx {\frac{1}{128}  
N_s} H_1^4 \biggl({z+1 \over z_1 +1}\biggr)^4 
\biggl({g_1 \over g_{eff}(z)}\biggr)^{1/3} ~~,
\label{rhobford}
\end{equation}
where $H_1$ and $z_1$ are the expansion factor and redshift at the 
matter thermalization epoch at the end of quintessential inflation, 
respectively.  The factor of 128 in this expression comes from the
explicit integration of the particle creation terms. 

When the gravitons and quanta of the  $Q$ 
field are formed  at the end of inflation,
one expects \cite{vilenkin}
the energy density in gravity waves  to be twice
the energy density in the $Q$-field quanta (because there are two graviton
polarization states).  In this paradigm then, wherever we have deduced a constraint
on $\rho_Q$, it should be taken as the sum of three different contributions.
One is the dark energy from the vacuum expectation value $\langle \rho_Q \rangle$
 of the $Q$ field; a second is the contribution from the
fluctuating part $\rho_{\delta Q}$ of the $Q$ field; and a third is
from the energy density $\rho_{GW}$ in relic gravity waves.  Thus, we have
\begin{equation}
 \rho_Q \rightarrow (\langle \rho_Q \rangle + \rho_{\delta Q} + \rho_{GW})~~.
\end{equation}
 For the cases of interest,
the energy density in gravity waves and quanta scales like radiation after
inflation, $\rho_{GW} + \rho_{\delta Q} \propto a^{-4}$, 
while  the quintessence field vacuum expectation value
evolves according to Eq.~(\ref{qdecay}) which gives
$\langle \rho_Q \rangle \propto a^{-6}$  during the kinetic dominated epoch. 
This epoch following inflation lasts until the energy in the $Q$ field
falls below the energy in background radiation and matter,
$\rho_Q \le \rho_B$.

Thus, for the kinetic dominated initial conditions (curve A)
in particular, gravity waves could be an important contributor
to the excess energy density during nucleosynthesis.
The relative contribution of gravity waves and quintessence quanta
compared to the background matter fields is just given by the
relative number of degrees of freedom.  At the end of inflation, 
the relative fraction of energy density in quanta and gravity
waves is given by  \cite{vilenkin}
\begin{equation}
(\rho_{\delta Q} + \rho_{GW})/\rho_B = 3/N_s~~,
\end{equation}
where $N_s$ is the number of ordinary scalar fields at the
end of inflation.  In the minimal supersymmetric
model $N_s = 104$.   
Propagating this ratio to the time of nucleosynthesis requires another
factor of  $(g_n/g_{1})^{1/3}$ where $g_n = 10.75$ counts the number of
effective
relativistic degrees of freedom just before electron-positron annihilation,
and $g_{1}$ counts the number of degrees of freedom 
during matter thermalization after the end of inflation.
In the minimal standard model  $g_{1} = 106.75$, but in supergravity
models this increases $\sim 10^3$.

 Combining these factors we have
\begin{equation}
(\rho_{\delta Q} + \rho_{GW})/\rho_B \le 0.014~~,
\end{equation}
during nucleosynthesis.  Hence, in this paradigm, the allowed values
of $\rho_Q/\rho_B$ consistent with nucleosynthesis 
could be reduced from a maximum of 0.056 to 0.042,
further tightening the constraints deduced here.

\subsection{Gravity-Wave Spectrum}

There has been considerable recent interest
\cite{riazuelo,giovannini1,giovannini2} 
in the spectrum of gravity waves
produced in the quintessential inflation paradigm.
One might expect that the {\it COBE} constraints on the spectrum 
also lead to constraints on the $Q$ field.  Here we consider this
possibility, but conclude that no significant constraint on the
initial $\rho_Q$ or effective potential is
derived from the gravity wave spectrum.  On the other hand,
the BBN and CMB gravity-wave  constraints discussed here 
can be used to provide useful constraints on the quintessential
inflation epoch as we now describe.
Our argument is as follows:
The  logarithmic gravity-wave energy  spectrum observed
at the present time can be defined in terms 
of a differential closure fraction,
\begin{equation}
\Omega_{GW}(\nu) \equiv {1 \over  \rho_c} {d \rho_{GW} \over d \ln{\nu}}~~,
\end{equation}
where the  $\rho_{GW}$ is the present energy density in relic gravitons 
and $\rho_c(0) = 3 H_0^2/8 \pi G = H_0^2 m_{pl}^2$ is the critical density. 
This spectrum  has been derived       
in several recent papers \cite{riazuelo,giovannini1,giovannini2}.  
It is characterized
by spikes at low and high frequency.  The most
stringent constraint at present derives from 
the {\it COBE} limit on the tensor component
of the CMB power spectrum  at low multipoles.  There is also a weak constraint from
the binary pulsar \cite{giovannini1,giovannini2} and an integral constraint
 from nucleosynthesis as mentioned above.

 For our purposes, the only possible
new constraint comes from the {\it COBE} limits
on the tensor component of the CMB power spectrum.  
The soft branch in the gravity-wave spectrum lies in the frequency range
between the present horizon 
$\nu_0 = 1.1 \times 10^{-18} \Omega_{M}^{1/2}h $ 
Hz and the decoupling frequency $\nu_{dec}(0) = 1.65 \times 10^{-16} 
\Omega_{M}^{1/2} h $  Hz,
where  we adopt $\Omega_{M} = 0.3$ for the  present matter
closure fraction.
The  constraint on the spectrum  can be
written \cite{giovannini2},
\begin{eqnarray}
\Omega_{GW}(\nu) &=& \Omega_\gamma  {81 \over (16 \pi)^2} \biggl(
{g_{dec} \over g_{1}}\biggr)^{1/3}
\biggl({H_1 \over m_{pl}}\biggr)^2 
\times \biggl({\nu_{dec} \over \nu}\biggr)^2 
\nonumber \\
&\times& 
\ln^2{\biggl({\nu_{r} \over \nu_1}\biggr)} \le  6.9 h^{-2} \times 10^{-11} ~~,
\label{gwspec}
\end{eqnarray}
where, $\Omega_\gamma = 2.6 \times 10^{-5} h^2$ is the present closure 
fraction in photons.  The number of relativistic degrees of freedom at
decoupling is $g_{dec} = 3.36$.  As noted previously,
$g_1$ is the number of relativistic degrees of
freedom after matter thermalization.  In the minimal
standard model is $g_{1} = 106.75$. The quantity $H_1$ is the expansion rate at 
the end of inflation.  In quintessential inflation it is  simply
related to the kinetic energy $\rho_Q$ after inflation, 
\begin{equation}
\rho_{Q}(z_1) = H_1^2 m_{pl}^2~~.
\end{equation}
The logarithmic term in Eq.~(\ref{gwspec}) involves  
present values of the  frequency $\nu_r(0)$
characteristic of the start of radiation domination ($\rho_B = \rho_Q$
at $z = z_r$), and 
the frequency characteristic of matter thermalization
at the end of inflation $\nu_1(0)$.
The quantity $\nu_1$ is just,
\begin{equation}
\nu_1(0) = {H_1 \over c}\biggl({z_0 + 1 \over z_1 + 1}\biggr)~~.
\end{equation}
Similarly,
\begin{equation}
\nu_r(0) = {H_r \over c}\biggl({z_0 + 1 \over z_r + 1}\biggr) = {H_1 \over c}\biggl({z_r + 1 \over z_1 + 1}\biggr)^3
\biggl({z_0 + 1 \over z_r + 1}\biggr)~~,
\end{equation}
so that,
\begin{equation}
{\nu_r(0) \over \nu_1(0)} = \biggl({z_r + 1 \over z_1 + 1}\biggr)^2~~.
\end{equation}
The identity $\rho_B = \rho_Q$ at $z = z_r$ then gives,
\begin{equation}
{\nu_r \over \nu_1} =  {N_s\over 128} 
\biggl({H_1 \over m_{pl}}\biggr)^2 \biggl({g_1 \over g_r}\biggr)^{1/3} ~~,
\end{equation}
where for the cases of interest $g_r = 10.75$.

Collecting these terms, we can then use Eq.~(\ref{gwspec}) 
to deduce a constraint on the expansion factor
$H_1$ at the end of inflation
\begin{equation}
H_1^2 < 1.4  \times 10^{-11} m_{pl}^2~~.
\label{h1lim}
\end{equation}
For kinetic dominated models, $\rho_Q(z_1)$ at the end of inflation
is simply related to the
energy density $\rho_Q$ at $z $,
\begin{equation}
\rho_{Q}(z)  = \rho_{Q}(z_1) \biggl({z+1 \over z_1+1}\biggr)^6~~.
\label{rhoqz}
\end{equation}
Similarly the background matter energy density scales as
\begin{equation}
\rho_{B}(z) = \rho_{B}(z_1) \biggl({g_{1} \over
g_{eff}(z)}\biggr)^{1/3} \biggl({z +1 \over z_1+1}\biggr)^4 ~~.
\label{rhobz}
\end{equation}

Considering the present energy density in photons and neutrinos, we can
find a relation between $H_1$ and $z_1$:
\begin{eqnarray}
\rho_{\gamma 0} + \rho_{\nu 0} &=& 1.1 \times 10^{-125} m_{pl}^4 \nonumber \\
 &=& {N_s \over 128} H_1^4 
\biggl({g_1 \over g_{dec}}\biggr)^{1/3} (z_1 + 1)^{-4},  
\label{rhonow}
\end{eqnarray}
We then deduce from this and  Eq.~(\ref{h1lim}) that $z_1 < 8.4 \times 10^{25}$.
These equations then imply that there is only a lower
limit on $\rho_Q/\rho_B$ given by the constraint on $H_1$.
Combining Eqs.~(\ref{rhobford}), (\ref{rhoqz}), and (\ref{rhobz}), we have
\begin{equation}
\rho_Q(z)/\rho_B(z) = {128 {m_{pl}}^2 \over N_s H_1^2} 
\biggl({g_{1} \over g_{eff}(z)}\biggr)^{-1/3} 
\biggl({z +1 \over z_1+1}\biggr)^2~~.
\end{equation}
At our initial epoch $z = 10^{12}$, we then deduce that , $\rho_Q(z)/\rho_B(z)
~^>_\sim  5.6 \times 10^{-18}$.  
This limit is not particularly useful because $\rho_Q$
field must exceed $\rho_B$ at $z_1$ in order
for the gravitational particle production paradigm to work.  
 The implication is  then
that all initial conditions in which the kinetic term dominates over
the background energy at $z_1$ 
are allowed in the quintessential inflation scenario.
Hence, we conclude that the 
gravity-wave spectrum does not presently constrain
the initial $\rho_Q$ or $V(Q)$ in the quintessential inflation model.

  Before leaving this discussion on quintessential inflation,
however, we remark
that the limits on $\rho_Q/\rho_B \le 560$ derived from the BBN
constraints discussed below, can be used to place a lower limit on the
expansion rate at the end of inflation in this model.  This in turn can 
be used to deduce a lower limit on the redshift for the end of 
quintessential inflation from Eq.~(\ref{rhonow}).  Thus, we have
\begin{equation}
2.2 \times 10^{-21} < {H_1^2 \over m_{pl}^2} < 1.45 \times 10^{-11} ~~,
\end{equation}
and in the minimal supersymmetric model
\begin{equation}
1.0 \times 10^{21} < z_1 < 8.4 \times 10^{25}~~.
\end{equation}
This nontrivial constraint then implies that the kinetic driven
quintessential inflation must 
end at an energy scale somewhere between about 10$^{8}$ and 10$^{13}$ GeV,
well below the Planck scale.   By similar reasoning one can apply
this argument to the gravity-wave spectrum from normal inflation as
given in \cite{giovannini2}.  We deduce an upper limit to the
epoch of matter thermalization of
$H_1 \le 3.1 \times 10^{-10} m_{pl}^2$ which implies
 $z_1 \le 7.3 \times 10^{28} [g(z_1)/3.36]^{1/12}$.
In this case there is no lower limit from BBN as there is no 
$Q$ field present after inflation.

\section{Results and Discussion}

The equations of motion [Eqs.~(\ref{qddot}) and (\ref{friedmann})] were evolved
for a variety of initial $Q$ field strengths and power-law parameters $\alpha$.
As initial conditions, the quintessence field was assumed 
to begin with equipartition,
i.e.~$\dot Q^2/2  = V(Q)$. This implies $w_Q = 0$ initially.  
This seems a natural and not particularly restrictive choice, since
$w_Q$ quickly evolves toward  the kinetic ($w_Q = +1$)
 or the tracker solution  [Eq.~(\ref{wq})] depending
upon whether one begins above or below the tracker curve.

Nevertheless, this initial condition does not encompass all possible cases.
For example, one might imagine a somewhat strange initial condition in which
$Q$ begins at a very large value such that $V(Q) << \rho_c(0)$
while $\rho_Q = \dot Q^2/2$ is arbitrarily large.
In this case, $\rho_Q$ will quickly decay away as $a^{-6}$ before
the present epoch 
and insufficient dark energy will be present at the current epoch.
Hence, it is possible that even though one starts with a large   
$\rho_Q$, one may not satisfy the present equation of state constraint.
We do not consider this as a likely case, however.  The natural initial
values for the quintessence field are
$Q \le m_{pl}$, so that sufficient dark energy should 
always be present for this potential choice.  Besides, postulating
an initially small value for $Q$ reintroduces the small $\Lambda$
problem which is what these models were invented to avoid.

\begin{figure}
\psbox[xsize=6.0in, rotate=r]{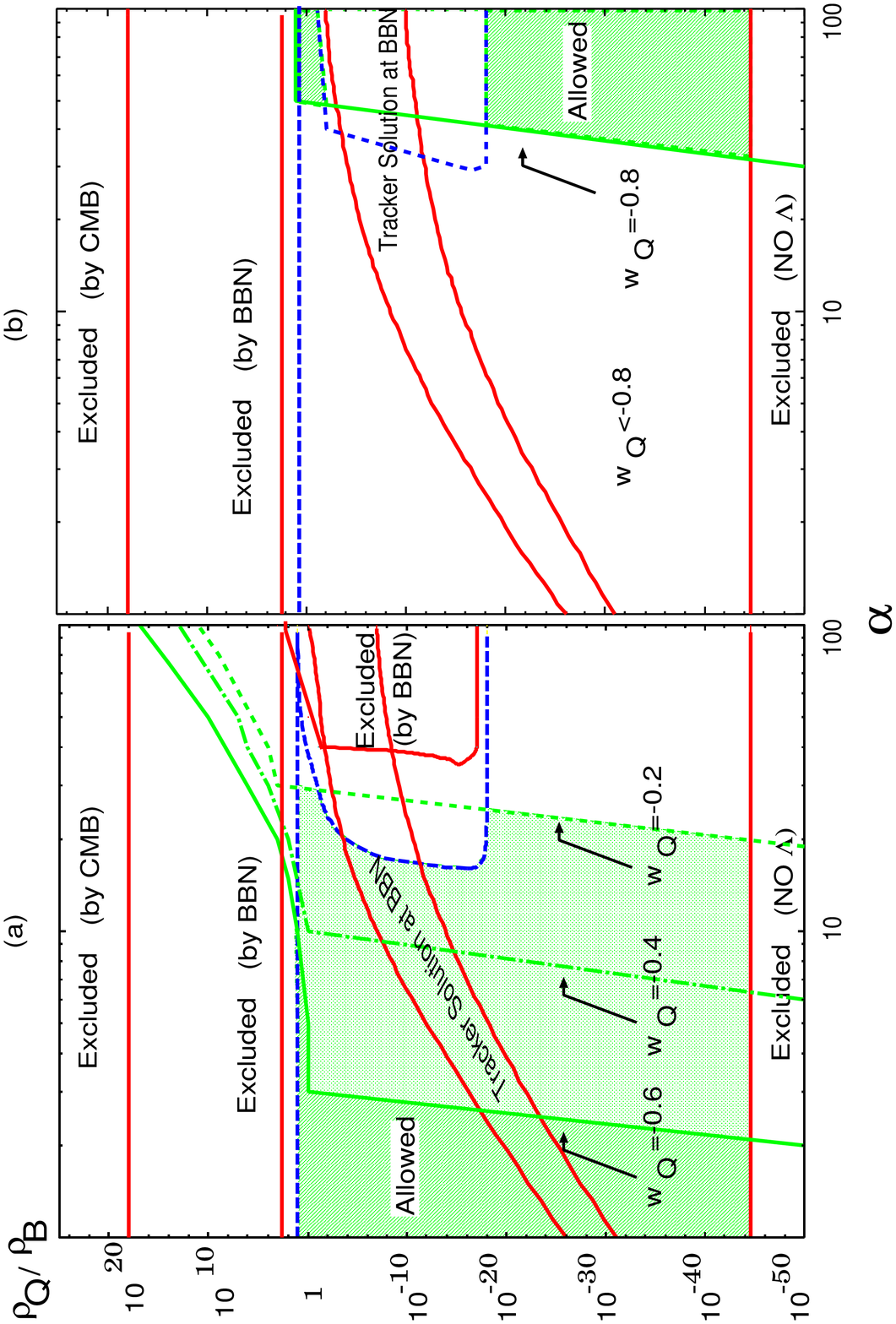}
\caption{
Contours of allowed values for $\alpha$ and initial $\rho_Q/\rho_B$ (at $z = 10^{12}$)
from various constraints as indicated for (a) the bare power-law
potential, and (b) the SUGRA corrected potential.
Models in which the tracker solution is obtained by the BBN epoch are 
indicated by the upper and lower curves. 
Values of $\alpha$ to the right of the lines labeled $w_Q = -0.6,-0.4,-0.2$ 
on (a) are excluded by the requirement that the present equation of
state be sufficiently negative.
The BBN constraint for a maximum energy density in the
$Q$ field of 0.1\% (dotted line)  and 5.6\% (solid line) are also indicated.
 For the SUGRA potential
(b) all tracker solutions to the right of the
region labeled $w_Q = -.8$ are allowed.}
\protect\label{fig:3}
\end{figure}

 Constraints on $\alpha$ and the initial value for the $Q$-field energy density
$\rho_Q(z)/\rho_B(z)$ at $z = 10^{12}$ were deduced numerically.  These
 are summarized in Figure \ref{fig:3} for both: (a) the bare
Ratra power-law  potential;  and  (b) its SUGRA corrected form.
We present the quantity $\rho_Q/\rho_B$ as a more intuitive
measure of the relative amount of $Q$-field energy density
than to just give $\rho_Q$.
This quantity can be easily converted to $\Omega_Q = (\rho_Q/\rho_B)/
(1 + \rho_Q/\rho_B)$.  
For purposes of illustration, we have arbitrarily 
specified initial conditions at $z = 10^{12}$, corresponding to 
$ T \sim 10^{12}$ K, roughly just after the time of the QCD epoch.
At any time the energy density $\rho_{rel}(z)$ in relativistic particles 
is just
\begin{equation}
\rho_{rel}(z) = \rho_{\gamma 0} 
\biggl({3.36 \over g_{eff}(z)}\biggr)^{1/3} (z+ 1)^4~~,
\end{equation}
where $\rho_{\gamma 0} = 2.0 \times 10^{-51} $ GeV$^4$,
is the present energy density in microwave background photons, and
we take  $g_{eff}(z) = 10.75$ between $z = 10^{12}$
and the beginning of BBN just before 
electron-positron annihilation ($z \approx  10^{10}$).

 The envelope of models which obtain a  tracker solution by the
epoch of nucleosynthesis are indicated by upper and lower
curved lines in Figures \ref{fig:3}a and \ref{fig:3}b. 
The general features of these constraints 
are as follows.  If the initial energy density in the $Q$
field is too large,
the tracker solution is not reached by the time of BBN. 
The $Q$-field energy density can
then significantly exceed the background energy  during nucleosynthesis.
This situation
corresponds to the excluded regions on the top of
Figures \ref{fig:3}a and  \ref{fig:3}b.  

All solutions
consistent with the  primordial nucleosynthesis 
 constraints are
also consistent with our adopted CMB $Q$-energy
constraint
 as  also shown at the top of 
Figures  \ref{fig:3}a and \ref{fig:3}b. 
Similarly, if the initial energy in the $Q$ field is too small,
the universe does not become   dark-energy dominated
by the present time.  This
corresponds to the excluded  (no $\Lambda$)
region at the bottom of the figures.  

The excluded regions at the top and bottom of figures \ref{fig:3}a 
and \ref{fig:3}b can be easily understood analytically.
For example, the excluded (no $\Lambda$) region at the bottom
of these figures reflects the fact that if $\rho_Q$ is initially below
the value presently required by  $\Omega_\Lambda = 0.7$ [cf. Eq.~(\ref{rhoq0})]
 it can not
evolve toward a larger value.  Hence, 
$\rho_Q/\rho_B < 2.8 \times 10^{-44}$ is
ruled out for $h = 0.7$.  
Similarly, the ''Excluded by BBN`` region comes from requiring
that $\rho_Q(z_{BBN})/\rho_B(z_{BBN}) < 0.056$.
For this constraint we are only considering cases in which
the $Q$ field is approaching the tracker solution from above 
during nucleosynthesis.  Hence, it is in the kinetic regime in
which $\rho_Q \propto z^6$ while the background scales as $z^4$.  
Thus, we have
 $\rho_Q(z=10^{12})/\rho_B(z=10^{12}) > 0.056 (10^{12}/10^{10})^{6-4}
 = 560$ is excluded.  
By similar reasoning, the ''Excluded by CMB`` region is given
by $\rho_Q(z_{CMB})/\rho_B(z_{CMB}) > 0.64 $,
or $\rho_Q(z=10^{12})/\rho_B(z=10^{12}) >  0.64 (10.75/3.36)^{1/3} 
(10^{12}/10^{3})^{6-4} = 9.4 \times 10^{17}$.

For the bare Ratra  power-law potential (Fig.~3a)  the main constraint is simply
the requirement that the equation of state be sufficiently negative by the 
present time.  
The sensitivity of 
the allowed power-law exponent to the equation of state is indicated
by the $w_Q =$ -0.2, -0.4, and -0.6 lines on Figure \ref{fig:3}a.   
In the present $Q$-dominated
epoch, Eq.~(\ref{wq}) is no longer valid even in the tracker 
regime.  Thus, the lines of constant
$w_Q$ must be evaluated numerically using Eq.~(\ref{wqnum}).  
The slight slope to these curves comes from the fact that 
$V(Q)/\rho(Q)$ has not yet reached unity,
i.e.~there is still some small kinetic contribution to $\rho(Q)$ and
the amount of kinetic contribution depends upon $\alpha$.

For the bare Ratra power-law potential, tracker solutions
with $\alpha ^<_\sim 20$ are allowed if $w_Q \le -0.2$. 
The allowed values for $\alpha$ reduce to $< 9$ and $< 2$
if the more stringent -0.4 and -0.6 constraints are adopted.
If $\alpha$ is too small, say $\alpha \le 2$,  then the potential
parameter $M$ becomes a very small fraction of the Planck mass
[cf. Eq.~(\ref{mfix})], and the fine tuning problem is reintroduced.

 For models in which the tracker solution is obtained by the time of
BBN, nucleosynthesis only limits the potential parameters 
if the most conservative equation of state limit ($w_Q < -0.2$)
 and most stringent nucleosynthesis constraint 
($\rho_Q/\rho_B < 0.1\%$) are adapted.   On the other hand,
independently of the equation of state constraint,
nucleosynthesis 
limits a large family of possible kinetic-dominated 
solutions in which the $Q$-field
energy density  exceeds the background energy prior to and during the 
nucleosynthesis epoch.  Such models are excluded
even though they provide sufficient   dark energy
in the present epoch.

For the SUGRA-corrected  $Q$ fields (Fig. 3b), 
the constraint from $w_Q$ is greatly relaxed.  In fact,
$w_Q$ is sufficiently
negative ($w_Q < -0.6$) for all $\alpha < 10^4$.
The reason is that all tracker solutions have $w_Q \approx -1$.
This is because $w_Q$ decays much faster toward -1 for
the  SUGRA potential.
Also, the potential has a finite minimum which is 
equal to the present dark-energy density.  
The Q field quickly evolves to near the potential minimum
and has negligible kinetic energy by the present time.
Any potential which becomes flat at late times 
gives $w_Q \approx -1$ and the   dark energy looks 
like a cosmological constant $\Lambda$.
All SUGRA models which achieve the tracker solution also have
a small $\rho_Q$ during 
primordial nucleosynthesis.
Hence, there is no constraint from nucleosynthesis except for those
kinetic-dominated models in which the $Q$ field is still far above the tracker
solution during the nucleosynthesis epoch.

We do note, however, that if a lower limit of
$w_Q > -0.8$ is adopted for tracker solutions
 from \cite{steinhardt99},
then only a power law with $\alpha ^>_\sim 30$ is allowed as indicated
on Figure 3b.  This makes the SUGRA potential the preferred 
candidate for quintessence.
The large $\alpha$ implies values of $M$ in Eq. (\ref{mfix})
close to the Planck mass, thus avoiding any fine tuning problem.
However, this quintessence potential is constrainable by BBN
should the light-element constraints become sufficiently precise to 
limit $\rho_Q$ at the 0.1\% level.

\section{Conclusions}

We conclude that for both the bare Rata  inverse power-law potential 
and its  SUGRA-corrected  form, the main constraints for models
which achieve the tracker solution by the nucleosynthesis epoch is
from the requirement that the equation of state parameter becomes
 sufficiently
negative by the present epoch.  The main constraint from nucleosynthesis
is for models in which the $Q$ field has not yet obtained 
the tracker solution by the time of nucleosynthesis. 
The SUGRA-corrected potential is the least constrained and
avoids the fine tuning problem for $M$.  Therefore,
it may be the preferred candidate for  the quintessence field,
although BBN may eventually limit this possibility.
We also note that the constraints considered here provide useful
constraints on the regime of matter creation at the end of quintessential 
inflation.

\acknowledgments
One of the authors (GJM) wishes to acknowledge the hospitality
of the National Astronomical Observatory of Japan where much of this
work was done.
This work has been supported in part by the Grant-in-Aid for
Scientific Research (10640236, 10044103, 11127220, 12047233) 
of the Ministry of Education, Science, Sports, and Culture of Japan, 
and also in part by 
DoE Nuclear Theory Grant (DE-FG02-95-ER40934 at UND).

%
%
%
%

%
\onecolumn
\widetext

\vfill\eject

\vfill\eject

%
%


\begin{references}
%

\bibitem{garnavich} P.~M.  Garnavich, et al.,  Astrophys. J., {\bf 509},
 74 (1998); Riess, et al., Astron. J, {\bf 116}, 1009 (1998). 
%
\bibitem{perlmutter} S. Perlmutter, et al., Nature {\bf 391}, 51 (1998);
Astrophys. J., {\bf 517}, 565 (1998). 
%
\bibitem{wang}  
L. Wang, R.R. Caldwell, J.P. Ostriker and P. J. Steinhardt, Astrophys. J., 
530, 17 (2000).   
%
\bibitem{wetterich}  
C. Wetterich, Nucl. Phys., B302, 668 (1988).
%
\bibitem{zlatev}  
I. Zlatev, L. Wang, and P. J. Steinhardt, Phys. Rev. Lett., 82, 896 (1999).
%
\bibitem{chiba}  
T. Chiba, T. Okabe, and M. Yamaguchi, Phys. Rev. D62, 023511, (2000).
%
\bibitem{armendariz1}  
C. Armendariz-Picon, V. Mukhanov, and P. J. Steinhardt, Phys. Rev Lett., {\bf 85},
4438 (2000).
%
\bibitem{armendariz2}  
C. Armendariz-Picon, V. Mukhanov, and P. J. Steinhardt, Phys. Rev D63, 103510
(2001)
%
\bibitem{steinhardt99}  
P.  J. Steinhardt, L. Wang and I. Zlatev, Phys. Rev. D59, 123504 (1999).
%
\bibitem{brax}  
P. Brax, J. Martin, and A. Riazuelo,  Phys. Rev D62, 103505 (2000).
%
\bibitem{vilenkin}  
P.J.E.~Peebles and A. Vilenkin, Phys. Rev. D59, 063505 (1999).
%
\bibitem{kolda99}  
C. Kolda and D. H. Lyth, Phys. Lett., {\bf B458}, 197 (1999);
S. Weinberg, in Proc. Dark Matter 2000, Marina Del Rey, Feb, 2000, 
astro-ph/0005265, (2000).
%
\bibitem{copeland}  
E. J. Copeland, N. J. Nunes, and F. Rosati, Phys. Rev. {D62}, 123503 (2000).
%
\bibitem{hellerman}  
T. Banks, JHEP submitted (2000), hep-th/0007146; 
T. Banks  and W. Fischler, JHEP submitted (2000), hep-th/0102077; 
S. Hellerman, N. Kaloper, and L. Susskind, JHEP, 06, 003 (2001);
W. Fischler et al., JHEP, 07, 003 (2001) hep-th/0104181;
C. Kolda and W. Lahneman, hep-th/0105300.
%
\bibitem{chenlin} B. Chen and F-L. Lin,  hep-th/0106054.
%
\bibitem{bean}  
R. Bean, S. H. Hansen, and A. Melchiorri, astro-ph/0104162, in press (2001).
%
\bibitem{as}  
A. Albrecht and C. Skordis, Phys. Rev. Lett., {\bf 84}, 2076 (2000).
%
\bibitem{chen}  
X. Chen, R. J. Scherrer, and G. Steigman, Phys. Rev D63, 123504 (2001).
%
\bibitem{ratra}  B.~Ratra and P.J.E.~Peebles, Phys.~Rev.~D37, 3406 (1988);
P.J.E.~Peebles and B.~Ratra, Astrophys.~J.~Lett., 325, L17 (1988).
%
\bibitem{sugraref}  
P. Brax and J. Martin, Phys. Lett., {\bf B468}, 40 (1999); Phys.Rev. D61 (2000) 103502;
P. Brax, J. Martin, A. Riazuelo, Phys. Rev. {\bf D62}, 103505 (2000).
%
\bibitem{osw99} K.A.~Olive, G.~Steigman, and T.P.Walker, T. P. Phys. Rep. 333,
389 (2000).
%
\bibitem{nollett00} K.M.~Nollett and S.~Burles, Phys. Rev. {\bf D61},
123505 (2000); S.~Burles, K.M.~Nollett, \& M.S.~Turner, Phys. Rev. {\bf D63},
063512 (2001).
%
\bibitem{tytler00} D.~Tytler, J.~M.~O'Meara, N.~Suzuki, \& D.~Lubin
Phys. Rep., 333, 409 (2000), astro-ph/0001318.
%
\bibitem{steigman00}  Steigman, G, in ``The Light Elements
and Their Evolution,'' IAU Symp. 198, L. ds Silva, M. Spite,
J. R. De Medeiros, eds., (ASP; San Francisco; 2000), pp. 13-24.
%
\bibitem{freese} K.~Freese, F.~Adams, J.~Frieman, and E.~Mottola,
in {\it Origin and Distribution of the Elements}, G.~J.~Mathews, ed.
(World Scientific: Singapore) (1988), pp. 97-115.
%
\bibitem{boom} C.B. Netterfield, et al. (Boomerang Collaboration),
Submitted to Astrophys. J. (2001), astro-ph/0104460.
%
\bibitem{dasi} C. Pryke, et al.
(DASI Collaboration), Submitted to Astrophys. J. (2001), astro-ph/0104490.
%
%
\bibitem{hu}  
W. Hu, D. Scott, N. Sugiyama, and M. White, Phys. Rev. D52, 5498 (1995).
%
\bibitem{hannestad}S.  Hannestad, Phys. Rev. Lett., 85, 4203 (2000).
%
\bibitem{riazuelo}  
A. Riazuelo, and J.-P. Uzan,   Phys. Rev. {\bf D62}, 083506 (2000).
%
\bibitem{giovannini1}  
M. Giovannini, Class. Quant. Grav {\bf 16}, 2905 (1999).
%
\bibitem{giovannini2}  
M. Giovannini, Phys. Rev. D60, 123511 (1999).
%
\bibitem{ford} L. H. Ford, Phys. Rev. {\bf D35}, 2955 (1987).
%
%
%
\end{references}
\end{document}